\documentstyle[aaspp4,epsf]{article} 

\slugcomment{Accepted on August 18, 1998 for publication by the 
{\it Astrophysical Journal (Letters)}}

\lefthead{Reach and Rho}
\righthead{Far-infrared H$_2$O and OH from SNR 3C~391} 

\begin{document} 
 
\title{Detection of Far-Infrared Water Vapor, 
 Hydroxyl, and Carbon Monoxide Emissions from the Supernova Remnant 3C~391
 \footnote{Based on observations with ISO, an ESA project with instruments funded by ESA Member States
 (especially the PI countries: France, Germany, the Netherlands and the
United Kingdom) with the participation of ISAS and NASA.}
 } 
 
\author{William T. Reach\altaffilmark{1}}
\affil{Infrared Processing and Analysis Center,
California Institute of Technology,
Pasadena, CA 91125}
\author{Jeonghee Rho\altaffilmark{2}}
\affil{Service d'Astrophysique, CEA, DSM, DAPNIA, Centre d'Etudes de Saclay,
F-91191 Gif-sur-Yvette cedex, France}

\altaffiltext{1}{formerly Institut d'Astrophysique Spatiale, B\^atiment 121, Universit\'e
Paris XI, 91405 Orsay cedex, France}
\altaffiltext{2}{postal address: Physics Department, University of California, 
Santa Barbara, CA 93106}

\def\etal{et al.}
\def\ISO{{\it ISO}}
\def\arcsec{\hbox{$^{\prime\prime}$}}
\def\kms{ {km~s$^{-1}$} }
\def\simgt{\mathrel{\mathpalette\oversim>}}
\def\simlt{\mathrel{\mathpalette\oversim<}}
 
\begin{abstract} 
We report the detection of shock-excited far-infrared 
emission of H$_2$O, OH, and CO from the supernova remnant
3C~391, using the {\it ISO} Long-Wavelength Spectrometer.
This is the first detection of thermal H$_2$O and 
OH emission from a supernova remnant. 
For two other remnants, W~28 and W~44, CO emission was detected
but OH was only detected in absorption.
The observed H$_2$O and OH emission lines arise from levels 
within $\sim 400$~K of the ground state, consistent with collisional excitation
in warm, dense gas created after
the passage of the shock front through the dense 
clumps in the pre-shock cloud. 
The post-shock gas we observe has a density $\sim 2\times 10^5$~cm$^{-3}$
and temperature 100-1000~K, and 
the relative abundances of CO:OH:H$_2$O in the emitting region are
100:1:7 for a temperature of 200 K. 
The presence of a significant column of warm H$_2$O 
suggests that the chemistry has been significantly changed by the shock.
The existence of significant column densities of both OH and H$_2$O,
which is at odds with models for non-dissociative shocks into dense gas,
could be due to photodissociation of H$_2$O or 
a mix of fast and slow shocks through regions with 
different pre-shock density.
\end{abstract} 
 
 
\section{Introduction} 
 
Long suspected to be an important component of the interstellar
medium, water has proven difficult to observe because of the
dominating absorption by water vapor in the Earth's atmosphere.
Water masers at radio frequencies are bright enough to penetrate 
the atmosphere, but
they are rare and their physical conditions may be exceptional.
Now that we have flown a sensitive infrared spectrometer in space,
aboard the {\it Infrared Space Observatory} (\cite{kessler}), 
transitions among energy levels $\sim 100$--1000 K above the
ground state of H$_2$O are observable. 
Early results from {\it ISO} indicate that H$_2$O is a
significant constituent in a cloud near the galactic center (\cite{cernicharo})
and that H$_2$O is as abundant as CO in shocked regions in Orion 
(\cite{harwit98}).

The observations reported here are part of a study of the 
infrared emission from supernova remnants interacting with molecular
clouds. Our targets were selected from a sample of supernova remnants with
previous evidence for interaction with nearby molecular clouds based on
X-ray and radio morphology (\cite{RP98}), millimeter-wave
molecular line observations (\cite{wilner98,woot77,woot81,rr98}), 
and OH 1720 MHz observations (\cite{Frail96,Green97}).
The presence of 1720 MHz maser emission without bright main-line 
OH maser emission suggests that they are
collisionally excited (\cite{elitzur}), and their location 
in bright radio remnants indicates that they 
are produced in the dense regions just behind molecular shocks.
The first results of our {\it ISO} observations showed that the
[O~I] 63 $\mu$m line, which is expected to be one of the brightest cooling
lines for post-shock gas for a wide range of gas densities and shock velocities,
is very bright at the locations of the OH masers (\cite{rr96}; Paper I).
In a parallel observational project, we also found shock-accelerated
CS, CO, and HCO$^+$ molecules in 3C~391 (\cite{rr98}).
In this paper, we present follow-up spectral observations to see which
additional infrared lines are bright from molecular supernova shocks.
These observations are the first detection of far-infrared H$_2$O,
OH, and CO lines from supernova remnants.
We report here far-infrared spectral observations of a single position in 
each of 3 supernova remnants: 3C~391, W~44, and W~28. 
 
\section{Observations} 
 
All observations reported here were performed with the 
{\it ISO} Long-Wavelength Spectrometer (\cite{clegg}).
We used the medium-resolution grating to fully sample the wavelength 
range from 42.2--188.6 $\mu$m. The coordinates are based on bright
OH 1720 MHz masers (\cite{Frail96}) in each remnant (for W~44:
18$^h$56$^m$28.4$^s$ +01$^\circ$29$^\prime$59$^{\prime\prime}$, for W~28: 
18$^h$01$^m$52.3$^s$ -23$^\circ$19$^\prime$25$^{\prime\prime}$; B1950); 
for 3C~391 we shifted closer to the peak of shocked 
CS and CO broad-molecular-line emission 3C~391:BML 
(18$^h$46$^m$47.1$^s$ -01$^\circ$00$^\prime$51$^{\prime\prime}$)
(\cite{rr98}).
The LWS beamsize is $80^{\prime\prime}$, and 
the spectra are severely affected by 
fringes, due to constructive and destructive interference (inside the
spectrometer) of wavefronts from our structured, extended sources.
We removed the fringes using
the {\it ISO} Spectral Analysis Package
\footnote{The ISO Spectral Analysis Package (ISAP) is a joint development by the 
LWS and SWS Instrument Teams and Data Centres.
Contributing institutes are CESR, IAS, IPAC, MPE, RAL and SRON.}, 
assuming that the emitting
region is extended at all wavelengths.
 
\begin{figure}
\plotone{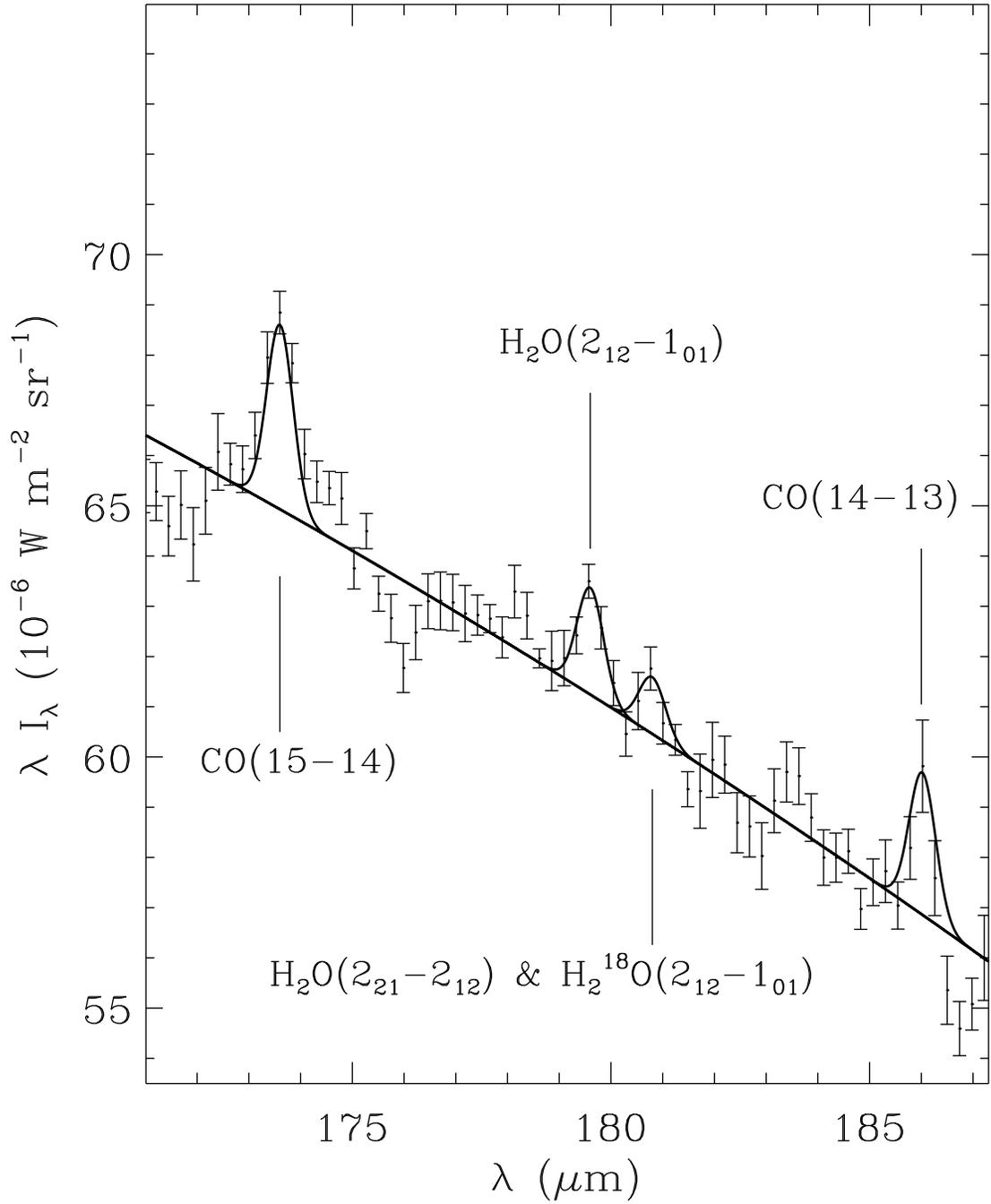}
\caption{A portion of the far-infrared spectrum of 3C~391:BML obtained the {\it ISO} LWS.
The combination of a linear baseline and 4 lines (with the instrumental
line profile) is shown as a solid curve.
Two H$_2$O lines and two CO lines are labeled by species and transition. 
\label{lwspec}}
\end{figure}

\begin{figure}
\plotone{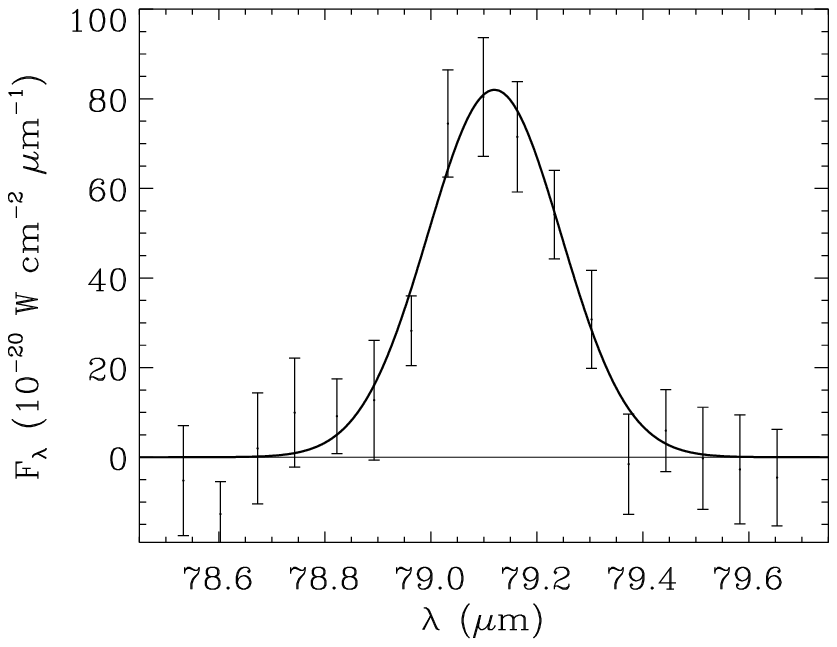}
\caption{Spectrum of the OH
[$^2\Pi_\frac{1}{2}(J=\frac{1}{2})-^2\Pi_\frac{3}{2}(J=\frac{3}{2})$]
line for 3C~391:BML.}
\label{ohspec}
\end{figure}


An important portion of the spectrum of 3C~391:BML,
including the low-lying transition of ortho-H$_2$O at 179.5 $\mu$m,
is shown in Fig.~\ref{lwspec}.
The continuum is due to dust
from the supernova remnant (some 30\% of the total) 
and unrelated interstellar material along the line of sight (see Paper I).
In addition to the H$_2$O and CO lines indicated, there are 
some remaining ripples in the spectrum that do not fall at the
wavelength of predicted lines nor do they have the shape expected
for an unresolved spectral line. 
 A second spectral observation toward 3C~391:BML was performed in order to
confirm some of the spectral lines, and to
search for lines of H$_2$O and OH from energy levels higher than we had already
detected. 
The result for the  OH
($^2\Pi_{\frac{1}{2}-\frac{3}{2}},J=\frac{1}{2}-\frac{3}{2}$)  line is shown in
Figure~\ref{ohspec}. 

\section{Results} 
 
Very bright lines from atomic and ionic C,
N, and O, and weaker molecular lines were detected from all three supernova
remnants. We will compile and present
the complete line lists in a future paper.
Of the three sources that we observed, only 3C~391:BML had a rich
spectrum of molecular emission. A list of H$_2$O, OH, and CO lines detected 
or limited toward 3C~391:BML is shown in Table~\ref{linetab}.  
For W~28, the CO(16-15) and CO(15-14) emission lines were detected, and for
W~44 only the CO(16-15) line was detected; the CO lines are a factor of 2
and 4 fainter for W~28 and W~44 compared to 3C~391:BML.
The 95\% confidence upper limit on the H$_2$O($2_{12}$-$1_{01}$)  179.5 $\mu$m line for
W~44 and W~28 is half the brightness of the line detected toward 3C~391:BML.
Considering that 3C~391 also had the brightest [O~I] 63 $\mu$m
line and dust continuum, it is possible that both W~28 and W~44 could have similar 
shock-excited spectra, with the H$_2$O lines unfortunately just below our detection
limit. While no OH emission was detected for W~28 and W~44, the 119 $\mu$m
transition from the ground-state was detected in {\it absorption}, presumably
due to foreground gas.
One source of uncertainty is the identification of the lines.
Although the resolving power of these observations was less than 300, 
such that wavelengths cannot be measured
sufficiently accurately for conclusive identifications,
our main conclusion---namely, that H$_2$O, OH, and CO were
detected from  shock-excited gas in supernova remnants---is
strong. In the
time-honored tradition of spectroscopy, we have detected  multiple lines from
each species.

\begin{deluxetable}{lllllll}
\tablecolumns{4}
\footnotesize
\tablecaption{Molecular Lines from 3C~391:BML\label{linetab}}
\tablewidth{0pt}
\tablehead{
\colhead{transition\tablenotemark{a}} & \colhead{flux} &
\colhead{$\lambda$\tablenotemark{b}} & \colhead{SNR} \\
\colhead{} & \colhead{(nW~m$^{-2}$~sr$^{-1}$)} & \colhead{($\mu$m)} & \colhead{} }
\startdata
CO (14-13)       	 & $9.0\pm 2.2$ & $186.06$ & 2.9 \nl 
CO (15-14)\tablenotemark{c}	 & $28.8\pm 2.3$ & $173.62$ & 6.4 \nl 
CO (16-15)\tablenotemark{c}       	 & $10.0\pm 1.7$ & $162.86$ & 5.0 \nl 
CO (17-16)       	 & $10.3\pm 2.8$ & $153.51$ & 3.0 \nl 
CO (18-17)\tablenotemark{c}  & $<17$ & 144.78 & \nodata \nl 
CO (19-18)      	 & $<17$ & 137.20 & \nodata \nl 
CO (20-19)       	 & $<7$ & 130.37 & \nodata \nl
H$_2$O ($4_{13}$-$4_{04}$) & $<8$ & 187.11 & \nodata \nl 
H$_2^{18}$O ($2_{12}$-$1_{01}$)\tablenotemark{c,d} & $ 6.6\pm 1.6$ & $180.69$ & 3.1 \nl 
H$_2$O ($2_{12}$-$1_{01}$) & $10.0\pm 0.9$ & $179.59$ & 8.2 \nl 
H$_2$O ($3_{03}$-$2_{12}$)\tablenotemark{c} & $ 6.3\pm 3.0$ & $174.54$ & 2.9 \nl 
H$_2$O ($3_{13}$-$2_{02}$) & $<13$ & 138.53 & \nodata \nl
H$_2$O ($4_{23}$-$4_{14}$) & $<7$ & 132.41 & \nodata \nl
H$_2$O ($4_{04}$-$3_{13}$) & $<13$ & 125.35 & \nodata \nl
H$_2$O ($4_{14}$-$3_{03}$) & $<5$ & 113.54 & \nodata \nl
H$_2$O ($2_{21}$-$1_{10}$) & $<7$ & 108.07 & \nodata \nl
H$_2$O ($2_{20}$-$1_{11}$)\tablenotemark{c} & $ 7.1\pm 2.0$ & $100.47$ & 2.9 \nl 
H$_2$O ($5_{15}$-$4_{04}$) & $<6$ & 95.63 & \nodata \nl
H$_2$O ($3_{22}$-$2_{11}$) & $<7$ & 89.99 & \nodata \nl
H$_2$O ($3_{21}$-$2_{12}$) & $<8$ & 75.38 & \nodata \nl
OH (1-1;3-1)\tablenotemark{c}   & $4.2\pm 1.2$ & $163.65$ & 2.6 \nl
OH (3-3;5-3)  & $<7$ & 119.33 & \nodata \nl
OH (1-3;5-7)  & $<13$ & 115.22 & \nodata \nl
OH (1-1;5-3)  & $3.6\pm 1.4$ & $98.69$ & 2.6  \nl
OH (1-3;3-5)  & $<9$ & 96.34 & \nodata \nl
OH (3-3;7-5)  & $12.7\pm 3.3$ & $84.58$ & 3.1 \nl
OH (1-3;1-3)  & $17.4\pm 1.9$ & $79.11$ & 11.9 \nl
OH (3-3;9-7)  & $<13$ & 65.20 & \nodata \nl
OH (1-3;3-3)  & $<13$ & 53.30 & \nodata \nl
\tablenotetext{a}{CO transitions are labeled in the format ($a$-$b$) where $a$ and $b$
are the upper and lower rotational quantum number; OH transitions are labeled in the
format ($a$-$b$;$c$-$d$) indicating the transition from level $^2\Pi_{a/2},J={c/2}$ to
level $^2\Pi_{b/2},J={d/2}$.}
\tablenotetext{b}{observed line center for detected lines, 
accurate to 0.2 (0.1) $\mu$m for 
$\lambda$ greater (less) than 88~$\mu$m; laboratory rest wavelength for
undetected lines}
\tablenotetext{c}{line potentially blended with another}
\tablenotetext{d}{This line could be the H$_2$O ($2_{21}$-$2_{12}$) transition
instead, but the excitation model suggests this transition contributes less 
than 10\% of the observed brightness.}
\enddata
\end{deluxetable}

\begin{figure}
\plotone{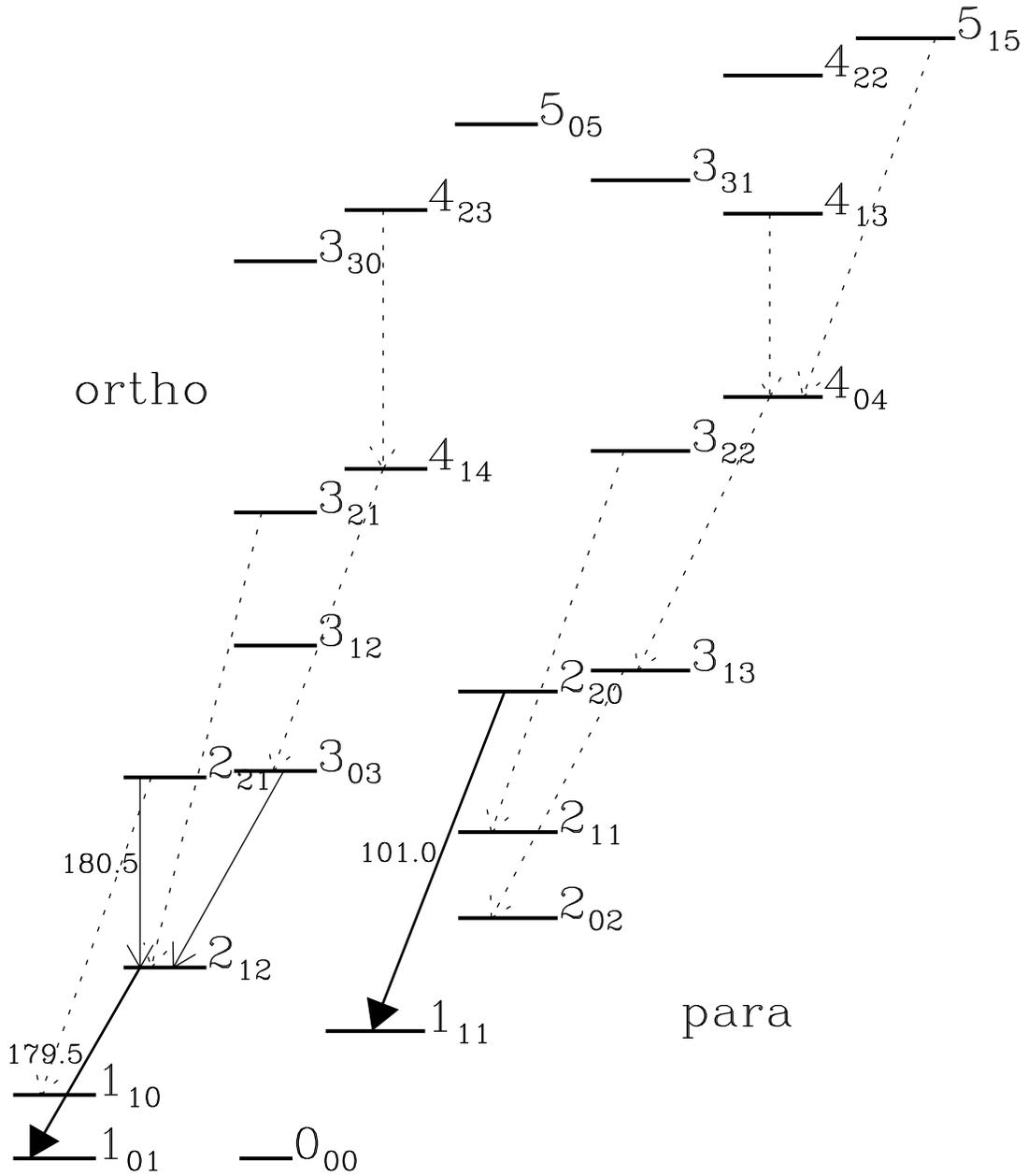}
\figcaption[f3.eps]{Energy-level diagrams for ortho (left) and para (right) 
H$_2$O. All levels  within 500~K of the ground state are shown.
The fastest radiative de-excitations of each level are
indicated with arrows, if the wavelength of the transition is
within the range of our {\it ISO} spectrum.
The arrows were drawn with the following characteristics: 
detected lines have a solid-filled arrowhead, 
possible  detections have an open arrowhead, 
and non-detections have dashed arrow. 
\label{waterlev} }
\end{figure}

\begin{figure}
\plotone{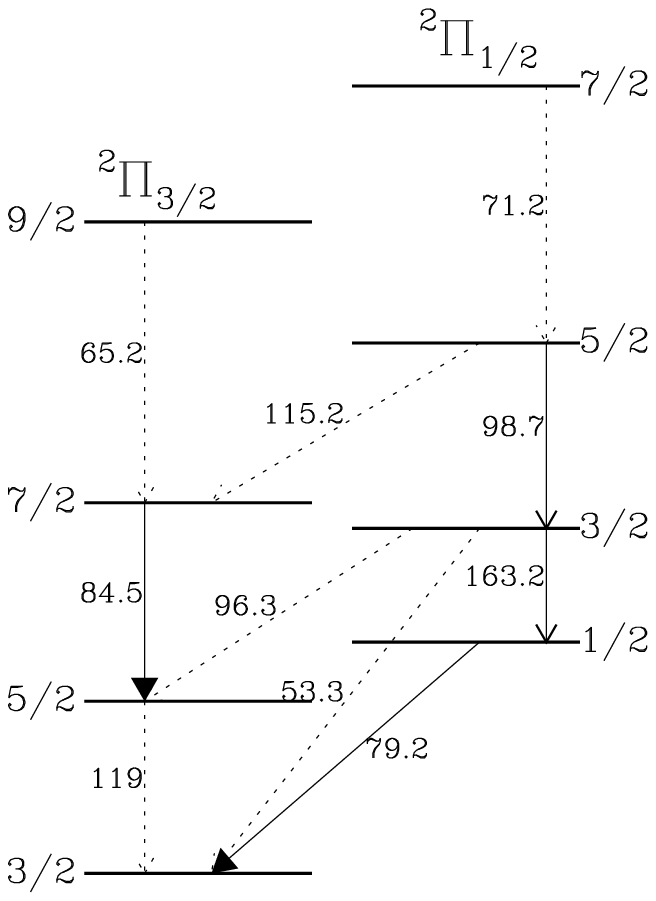}
\figcaption[f4.eps]{Energy-level diagram of OH. Each level
is labeled by its rotational quantum number $J$.
For the transitions within the range of our {\it ISO} 
spectrum, arrow symbols are the same as in Fig.~\ref{waterlev}.
Clearly detected lines have a solid-filled arrowhead, possible 
detections have an open arrowhead, and non-detections have 
dashed arrow. The transitions are labeled with the wavelength in $\mu$m.
\label{ohlev} }
\end{figure}

All of the lines we believe we have detected are from the lowest 
accessible energy levels of each species.
For CO, the highest level we detected was $J=16$, which lies 750~K
above the ground state. The energy level diagrams of H$_2$O and OH
are shown in Figures~\ref{waterlev} and~\ref{ohlev}.
In both cases, we indicate the fast transitions whose
wavelengths that fall within the range of our {\it ISO} spectrum. 
Including marginal detections,
we see evidence for almost all of the fast transitions of both ortho and para H$_2$O
among energy levels within 320~K of the ground state. 
The H$_2$O($4_{13}-3_{22}$) transition is masked by a bright [O~I] line nearby, but
there is a hint of the principal line from the next-highest energy level 
($3_{30}-2_{21}$).
No transitions involving levels more than 400~K above ground are seen.
The most important emission line that is expected to be bright but is
not detected is the 119 $\mu$m line of OH. But this line
connects directly to the ground state, and our models (below) predict 
significant foreground absorption; therefore, we suspect that the 119 $\mu$m
line is extinguished by foreground gas. Indeed, our spectra of
W~44 and W~28, which have comparable sightlines through the galactic disk,
show absorption features at 119 $\mu$m (with an optical
depth of order 0.1) due to foreground
OH absorbing the dust continuum from the supernova remnants.

\section{Discussion} 
 
\subsection{Excitation and abundance of H$_2$O, OH, and CO}

In order to determine the physical conditions that can produce
the observed suite of spectral lines, and to measure the abundances of
the observed species, we compared the line brightnesses to a simple
model that balances collisional and radiative transitions within a
uniform emitting region. 
We modeled the emission spectra of regions with a range of H$_2$ 
volume density and kinetic temperature, and the absorption spectrum of
a cold slab of foreground gas with nominal molecular abundances
(from \cite{irvine}) and an H$_2$ column density of $10^{22}$ cm$^{-2}$.
We solved iteratively for the excitation,
modifying radiative rates by the escape probability for a line
profile with a width of 30 km~s$^{-1}$, as was found from the
millimeter-wave CS and CO observations (\cite{rr98}).
Collision rates were taken from Offer, van Hemert, \& van Dishoeck (1994)
for OH and Green, Maluendes, \& McLean (1993) for H$_2$O, and
the radiative transition rates were taken from Pickett et al. (1996).
The observed brightness ratio of 79 $\mu$m to
84 $\mu$m lines of OH is sensitive to the gas density,
suggesting $n({\rm H}_2)\simeq (1-6)\times 10^5$ cm$^{-3}$.
The ratios of the $5\rightarrow 4$, $3\rightarrow 2$, and $2\rightarrow 1$
millimeter lines of CS are also sensitive to the gas
density, suggesting $n({\rm H}_2)\simeq (3-4)\times 10^5$ cm$^{-3}$,
and the millimeter CS and CO lines somewhat constrain the temperature, 
$T> 50$ K (\cite{rr98}).
For the abundance calculations, we will assume
$n({\rm H}_2)=2\times 10^5$ cm$^{-3}$ and $100<T<1000$ K,
which is consistent with the presence and lack of other OH and H$_2$O lines
in the observed wavelength range.

We determine the abundances of the various molecules 
assuming all lines arise from the same
physical region, with constant temperature and density.
The column density of OH is $\sim 2\times 10^{16}$ cm$^{-2}$, and the optical
depth of the 79 $\mu$m line is of order unity.
The CO excitation is much more sensitive to temperature; to match the 
brightness of the well-detected $15\rightarrow 14$ line, the
CO column density $\sim 2\times 10^{18} (T/200)^{-5}$ cm$^{-2}$.
The H$_2$O lines are estimated to be optically thick:
if we associate the line observed at 180.69 $\mu$m with 
H$_2^{18}$O ($2_{12}$-$1_{10}$), then the isotope ratio implies
an optical depth $\sim 400$ for the brightest H$_2$O line.
However, in the low-density limit, $n({\rm H}_2) < 10^9$ cm$^{-3}$,
spontaneous decay is still faster than collisional de-excitation, and
the line brightness still measures the H$_2$O abundance (\cite{irvine}).
The column density of H$_2$O from the excitation model is
$\sim 3\times 10^{17}$ cm$^{-2}$.
To determine the absolute abundance of each species, we require a measure of 
the H$_2$ column density.
If we assume the emitting region is a uniform
sphere with diameter equal to the observed angular size of shock-excited
molecular gas ($\sim 30^{\prime\prime}$ from \cite{rr98}), and we take
from the excitation model $n({\rm H}_2)=2\times 10^5$ cm$^{-3}$, we find 
$N({\rm H}_2)\sim 8\times 10^{23}$ cm$^{-3}$. 
This value of $N({\rm H}_2)$ agrees with the observed brightness 
(from our {\it ISO} SWS observations, in preparation)
of the S(3) line of H$_2$, if $T\simeq 200$.
In summary, we estimate the relative abundances of CO:OH:H$_2$O to be
100:1:7, and the abundance of water in the shocked cloud is 
[H$_2$O/H$_2$]$\sim 4\times 10^{-7}$.

\subsection{Chemistry of H$_2$O and OH}

The warm H$_2$O in 3C~391:BML is likely due to shock-enhanced chemistry. 
In relatively hot gas, OH is rapidly converted into H$_2$O 
by the reaction ${\rm OH} + {\rm H}_2 \rightarrow {\rm~H}_2{\rm O} + {\rm~H}$,
so that all the available O that is not already locked in CO
 would be converted into H$_2$O (\cite{DRD,MH}). 
Models for the oxygen chemistry predict very efficient conversion
of OH into H$_2$O for non-dissociative shocks into high-density 
($n_0=10^5$~cm$^{-3}$) clouds
but comparable OH and H$_2$O column densities for shocks into 
intermediate-density ($n_0=10^3$~cm$^{-3}$) clouds, while CO and H$_2$O
are comparably abundant (\cite{graff87}).
Behind a fully dissociative shock, 
the molecules re-form in a much cooler ($\sim 100$--500~K) region (\cite{HM89}).
In cooler gas, the 1420~K barrier (\cite{wg87}) for the
${\rm OH} + {\rm H}_2 \rightarrow {\rm~H}_2{\rm O} + {\rm~H}$
reaction cannot be
overcome, and OH is more abundant than H$_2$O (\cite{MH,vdB})
except deep inside dense molecular cores (\cite{sternberg}).
The chemistry in cooler gas depends on photo-dissociation;
OH is much more abundant than H$_2$O 
(\cite{vdB}), except deep inside very dark cores,
where the photodissociation rate is very low (\cite{bergin95}).

The relative strengths of the H$_2$O lines that we observe were calculated by
Kaufman \& Neufeld (1996), and our observations are generally
consistent with the models for pre-shock density $n_0=10^{4-5}$ cm$^{-3}$.
However, these models predict a very low OH abundance:
the H$_2$O 179.5 $\mu$m line is predicted to be
two orders of magnitude brighter than 
the OH 84.5 $\mu$m doublet, while we observe comparable brightnesses.
The relatively low excitation we observe for the H$_2$O and OH molecules ($T_{upper}<400$~K)
suggests that the gas we are observing is not presently hot enough for the 
rapid H$_2$O production. But this does not preclude the H$_2$O having formed
in a short-lived, high-temperature region, and we observe only the cooling region.
We observe {\it more} H$_2$O than OH, suggesting that high-temperature chemistry
was operative long enough to leave
a lasting effect on the chemistry of this gas.

The detection of OH, with an abundance only 7 times less than
that of H$_2$O, disagrees with theoretical models
for C shocks, which predict nearly
complete conversion of OH into H$_2$O.
This suggests either
(1) the H$_2$O abundance is underestimated due to beam-dilution,
(2) we are observing OH and H$_2$O from different types of shocks
in regions with a range of pre-shock densities, or 
(3) the models are not appropriate for the molecular shocks in 3C~391.
We detected bright ionic lines (including [O~III] and [N~III]) that are
only expected from dissociative shocks, so we know that a range of
shocks is present within our beam. 
However, the infrared OH line ratios are consistent with coming from
a region with the same physical conditions as inferred from 
the infrared H$_2$O, millimeter-wave CS, and radio-wave OH maser lines.
The disagreement between observations and predictions is at least
partially due to the fact that the models are for non-dissociative
shocks (\cite{KN96}). Including the effects of
photodissociation of H$_2$O by the interstellar radiation
field---as well as radiation local to the remnant---
could produce OH in the abundance we observe 
(Lockett, Gauthier, \& Elitzur 1998).

The water abundance we derive for 3C~391:BML 
is significantly lower than that observed from
molecular shocks in HH 54 (\cite{liseau}) or Orion
(\cite{harwit98}). In particular, in the Orion BN-KL region,
the H$_2$O abundance
is [H$_2$O/H$_2$]$\simeq 5\times 10^{-4}$
(\cite{harwit98}), incorporating nearly all of the oxygen in the gas.
The difference between the Orion BN-KL shock and the 3C~391:BML
shock could be due to limited angular resolution: 3C~391 is about 20 
times further away than Orion, while the angular resolution of our
observations is the same as that of Harwit et al. (1998). The difference
could also be due to details of the interaction between the
energetic events (a steady stellar wind in Orion {\it vs.}
impulsive supernova shock in 3C~391) and the interstellar 
medium (a molecular cloud and H~II region in Orion
{\it vs.} a giant molecular cloud with no H~II region in 3C~391). 

\section{Conclusions}

For the first time, thermal emission from the lower energy-levels of
H$_2$O and OH were detected from a supernova remnant. The emission arises
from gas cooling behind a shock front that is impinging
a particularly dense clump in the parent molecular cloud;
this site is called 3C 391:BML (broad molecular line).
Our spectra of interaction sites in W~44 and W~28
did not reveal OH or H$_2$O emission, although CO emission
and OH (foreground) absorption were detected.
3C~391 is not a unique case of a supernova remnant-molecular cloud 
interaction, but the infrared molecular lines imply the presence of
higher-density clumps than near W~44 and W~28, similar to IC~443.
Together with outflows from young stars, supernova remnants are
energetic events capable of altering cloud chemistry and producing
substantial columns of H$_2$O and OH.
Supernova-molecular cloud interactions are probably
common, and they have been suggested as an explanation for a new
class of mixed-morphology X-ray and radio supernova
remnants (\cite{RP98}). 
We can expect substantial advances in understanding shock-induced chemistry
in supernova-molecular cloud interactions by using future
observatories capable of higher angular and spectral resolution
in the far-infrared, such as planned for SOFIA (\cite{SOFIA}) 
and FIRST (\cite{FIRST}),
where we will better resolve the fast and slow shocks into gas
of varying pre-shock density, and we will better resolve a
larger number of spectral lines from each molecule.

\def\extra{
We have detected radiative transitions among levels of H$_2$O and OH
that can, in some excitation conditions, produce masers.
For H$_2$O, masers are predicted at 183 and 380 GHz,
originating from the $3_{13}$ and $4_{14}$ levels, respectively
(\cite{Neufeld91}). For OH, 1720 MHz maser line has already been detected
(\cite{Frail96}). The H$_2$O and OH significantly affect its
chemistry and cooling in the warm,
post-shock gas significantly affects its cooling and further chemistry.
The larger-scale effects on the thermodynamics and chemistry of interstellar
gas depend on how common are shocks that produce H$_2$O and OH,
and how long these molecules survive.
}
 
\acknowledgements 
The observations that we describe in this paper are part of the {\it ISO} open 
time granted to U.S. astronomers thanks to cooperation between 
the ESA and NASA. The research described in this paper was carried out in
part by the California Institute of Technology, under a contract with 
the National Aeronautics and Space Administration.

\end{document}